\documentclass[review]{elsarticle}
\usepackage{lineno,hyperref}
\usepackage{amsmath}
\pdfoutput=1
\modulolinenumbers[5]
\journal{Journal of \LaTeX\ Templates}
\usepackage{color}
\usepackage[dvipsnames]{xcolor}
\usepackage[normalem]{ulem}

\begin{document}
\begin{frontmatter}
\title{Topological properties of a chain of interacting electrons}
\tnotetext[mytitlenote]{Topological states in chain of interacting electrons}
\author{Igor N.Karnaukhov}
\address{G.V. Kurdyumov Institute for Metal Physics, 36 Vernadsky Boulevard, 03142 Kiev, Ukraine\\
Donostia International Physics Center (DIPC), 20018 Donostia/San Sebastián, Basque Country, Spain}
\author{E. E. Krasovskii}
\address{Donostia International Physics Center (DIPC), 20018 Donostia/San Sebastián, Basque Country, Spain \\
Departamento de Polímeros y Materiales Avanzados: Física,
Química y Tecnología, Universidad del Pais Vasco/Euskal Herriko Unibertsitatea,
20080 Donostia/San Sebastián, Basque Country, Spain \\
IKERBASQUE, Basque Foundation for Science, 48013 Bilbao, Basque Country, Spain}
\fntext[myfootnote]{karnaui@yahoo.com}
\begin{abstract}
Within the framework of a one-dimensional model of interacting electrons, the ground state of an electron liquid is studied. Using the exact solution of the model, the ground state phase diagram and zero-energy Majorana edge functions in a finite chain are calculated. The winding number invariant reflects the topological nature of the electron liquid. The phase diagram includes two topological phases with different winding number invariants, the topologically trivial Mott insulator phase, and three critical phase transition points. Numerical calculations confirm and illustrate the analytical results.
\end{abstract}
\begin{keyword}
\texttt  Kitaev chain \sep topological phase \sep zero energy Majorana edge mode
\end{keyword}
\end{frontmatter}

\section{Introduction}
Despite its simplicity, the Kitaev chain \cite{AK} is popular and physically demanded  because it describes topological superconductivity. The Hamiltonian of the model has a quadratic form, which allows us to study its properties in detail \cite{A1,A2,A3,A4,A5,AD1,AD2}. The presence of the interaction between electrons or a disorder can kill the topological phase, which raises an urgent question about its stability. It is clear that in the presence of the Hubbard interaction the band insulator becomes a Mott insulator (MI) in the limit of sufficiently large interaction. In 2D and 3D systems with strong interaction and half-filling, a
  Mott gap separating the lower and upper electronic bands is formed
  in the spectrum of charge excitations. This is the case in the Hubbard
  model; according to the Lieb-Wu solution~\cite{9}, an arbitrary value of
  the on-site interaction leads to an insulator state in the Hubbard
  chain. This gap has a many-particle nature, and the phase state at
  half-filling is close in nature to the Peierls phase. At the same
  time, this phase is not topological, and at a certain value of the
  on-site interaction one should expect a topological phase transition
  from the topological phase (if it is realized in the absence or weak
  interaction) to a non-topological state of the Mott insulator. The
question arises: At what value of the interaction does this phase
transition take place and what is its scenario?

Experimental studies of the topological states of the
  electron liquid in low-dimensional systems and cold atoms have
  inspired much theoretical~\cite{AD3} and experimental~\cite{AD4}
  effort. Usually, the topological state with an integer topological
  invariant is described in the framework of a one-particle band
  insulator model. However, when the interaction between the electrons
  is much stronger than the hopping integral the one-electron
  approximation is insufficient.  This paper is a step forward in
  solving this problem.
  
The list of exactly solvable models that account for the interactions in the Kitaev chain is short. These are, first of all, the Mattis-Nam (MN) model \cite{MN}, which describes the Kitaev chain for electrons with the on-site Hubbard interaction, the model for spinless fermions with the nearest neighbor interaction \cite{1}, and the MN chain, in which correlated hoppings between electrons located at the neighboring lattice sites are taken into account~\cite{IK}. The MN transformation allows an exact solution of the interacting model (without resorting to a mean field approximation). The Kitaev chain for electrons with the on-site Hubbard interaction $U(n_{j\uparrow}-\frac{1}{2})(n_{j+1 \downarrow}-\frac{1}{2})$ is exactly solvable for $\Delta=t$ and chemical potential $\mu=0$ \cite{MN}, where $t$ and $\Delta$ are the hopping integral and pairing of electrons located at the nearest neighbor lattice sites, respectively, and $n_{j \sigma}$ is the density operator of electrons, see the Hamiltonian (\ref{eq:1}).  The model is reduced to a spin-$\frac{1}{2}$ $XX$-chain in a magnetic field. The ground state phase diagram includes both a topologically nontrivial phase at $\vert U \vert <4 t$ and a topologically trivial one at $\vert U \vert >4t$.  The Kitaev chain for spinless fermions with the nearest neighbor interaction $U(n_j-\frac{1}{2})(n_{j+1}-\frac{1}{2})$ is also exactly solvable at the symmetry point $\Delta=t$ and $\mu=0$ \cite{1}.  The model is reduced to the spin-$\frac{1}{2}$ anisotropic $XY$ chain in zero magnetic field.  Thus, these models share the same simple phase diagram, namely, the topological phase is realized at $\vert U \vert <4 t$ \cite{MN} or at $\vert U \vert <2 t$ \cite{1}. The  model \cite{IK}, in which the phase diagram of the ground state has the same form, is not an exception. Obviously, the perturbation theory is inapplicable in this case, since the phase transition point from the topological  to the non-topological phase corresponds to a sufficiently strong interaction with a coupling of intermediate magnitude. Thus, the exact solution of the problem is the exact solution for the model Hamiltonian is required.

Let us pose the problem of how the topological properties of interacting electrons and correlated electron hopping between next-nearest neighbor lattice sites are related. To investigate this problem, we consider a one-dimensional (1D) model in which the topological state is realized and, at the same time, the on-site Hubbard interaction and correlated hoppings of electrons between next-nearest neighbor sites are taken into account (see Fig.~1). We take as a basis the MN model~\cite{MN} and add the correlation hoppings between electrons located at the next-nearest neighbor sites. The model is interesting and fruitful because it describes topological states of interacting electrons (the zero-energy Majorana edge modes) and has an exact solution. An exact solution exists even in the case of nondegenerate electron states, and it makes it possible to solve the 1D Falicov-Kimball model with $p$-pairing.

\section{The model}

The Hamiltonian of 1D model is the sum of two terms ${\cal H}={\cal H}_{MN}+{\cal H}_{1}$, the first of which is determined in accordance with the MN model, and the second takes into account the correlated hopping and pairing terms,
\begin{eqnarray}
&&{\cal H}_{MN}=
-\frac{1}{2}t\sum_{j}\sum_{\sigma}(c^\dagger_{j,\sigma}-c_{j,\sigma})( c^\dagger_{j+1,\sigma}+c_{j+1,\sigma})+ 
U\sum_{j}n_{j,\uparrow}n_{j,\downarrow}, \nonumber\\
&&{\cal H}_{1}=-2 v\sum_{j}\sum_{\sigma}(c^\dagger_{j-1,\sigma}-c_{j-1,\sigma}) (c^\dagger_{j+1,\sigma}+c_{j+1,\sigma})n_{j,-\sigma},
\label{eq:1}
\end{eqnarray}
where $c^\dagger_{j,\sigma},c_{j,\sigma}$ are the fermion operators defined on a lattice site $\emph{j}$ with spin $\sigma=\uparrow$ or $\downarrow$, $U$ is the on-site Hubbard interaction, $n_{j,\sigma}=c^\dagger_{j,\sigma}c_{j,\sigma}-\frac{1}{2}$, and 
$v$ determines the correlated hopping and pairing terms in (\ref{eq:1}) for fermions at the next-nearest neighbor sites. 

Using the Jordan-Wigner transformation, it is convenient to introduce spin-$\frac{1}{2}$ operators $\textbf{S}_j$ and $\textbf{T}_j$ defined via $c-$operators: $S^-_j=c_{j,\uparrow}(\prod_{i=1}^{j-1} S^z_{i})$, $T^-_j=c_{j,\downarrow2}(\prod_{i=1}^{j-1} T^z_{i})(\prod_{i=1}^{N} S^z_{i})$,   $S^z_j=n_{j,\uparrow}$, and $T^z_j=n_{j,\downarrow}$.
In the spin representation the Hamiltonian (\ref{eq:1}) has the following form
\begin{eqnarray}
&&{\cal H}_{MN}= -2 t\sum_{j}(S_{j}^x S_{j+1}^x +T_j^x T_{j+1}^x)+U\sum_{j=1}^{N}S_j^zT_j^z,\nonumber \\
&&{\cal H}_{1}= -8 v \sum_{j}(S^x_{j-1}S^x_{j+1} + T^x_{j-1}T^x_{j+1})T^z_j S^z_j. 
\label{eq:2}
\end{eqnarray}
The authors of \cite{MN} introduced a new set of spin-1/2 operators $\textbf {J}_j$ and $\textbf {P}_j$:
$\{S^x_j,S^y_j,S^z_j\} = \{J^x_j, 2J^y_jP^x_j, 2J^z_jP^x_j\}$ and $\{T^x_j,T^y_j,T^z_j\}= \{-2P^z_jJ^x_j, 2P^y_jJ^x_j, P^x_j\}$,
 which makes it possible to rewrite the general Hamiltonian in the following form
\begin{eqnarray}
&&{\cal H}=- 2t \sum_{j}J^x_jJ^x_{j+1}(1+4 P^z_j P^z_{j+1})+\frac{1}{2} U\sum_{j} J^{z}_{j}-\nonumber \\
&& 4 v\sum_{j}J^x_{j-1}J^z_j J^x_{j+1}(1 +4  P^z_{j-1} P^z_{j+1}).
\label{eq:3}
\end{eqnarray}
The $P^z_j$ operators commute with the Hamiltonian (\ref{eq:3}), i.e., they are the integrals of motion. The $P^z_j$ operators form a static $Z_2$ gauge field, and the uniform configuration of this field $P^z_{j}P^z_{j+1}=\frac{1}{4}$ corresponds to the ground state. It also follows from the numerical calculations that, although the $T$-spin chain is also determined by the $P^z_j$-operators, its behavior does not depend on their configuration. Thus, the symmetry between the $S-$ and $T-$spin chains is restored, since they both describe the spin degenerate states of electrons. 
Let us redetermine the Hamiltonian $\cal H$ (\ref{eq:3}) in terms of the operators of the spinless fermions $a^\dagger_j$ and $a_j$
\begin{eqnarray}
&&{\cal H}= -\frac{t}{2}\sum_{j}(a^\dagger_{j}-a_{j})( a^\dagger_{j+1}+a_{j+1})(1 +4 P^z_jP^z_{j+1}) +\frac{1}{2} U\sum_{j}\left(a^\dagger_j a_j - \frac{1}{2}\right)
\nonumber\\&&
-\frac{v}{4}\sum_{j}(a^\dagger_{j-1}-a_{j-1})( a^\dagger_{j+1}+a_{j+1})(1 +4 P^z_{j-1}P^z_{j+1}).
\label{eq:4}
\end{eqnarray}

\begin{figure}[tp]
     \centering{\leavevmode}
     \begin{minipage}[h]{.75\linewidth}
\center{
\includegraphics[width=\linewidth]{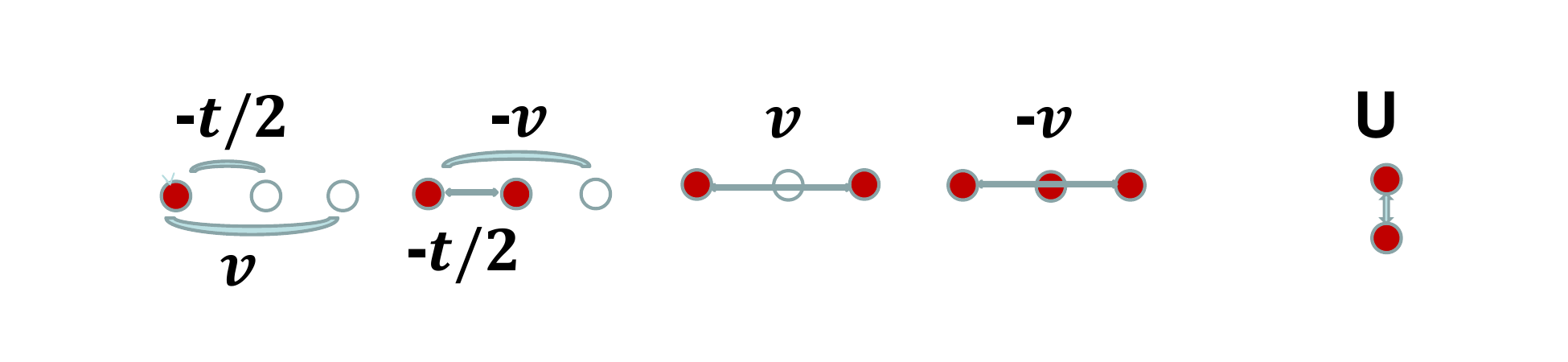}
                  }
    \end{minipage} \hfill
\caption{(Color online)
  A model of electron hopping and pairing along the chain between
  nearest-neighbor $-\frac{t}{2}$ and next-nearest-neighbor sites $\pm v$.
  $U$ is the on-site interaction.
}
\label{fig:0}
\end{figure}

\begin{figure}[tp]
     \centering{\leavevmode}
    \begin{minipage}[h]{.65\linewidth}
\center{
\includegraphics[width=\linewidth]{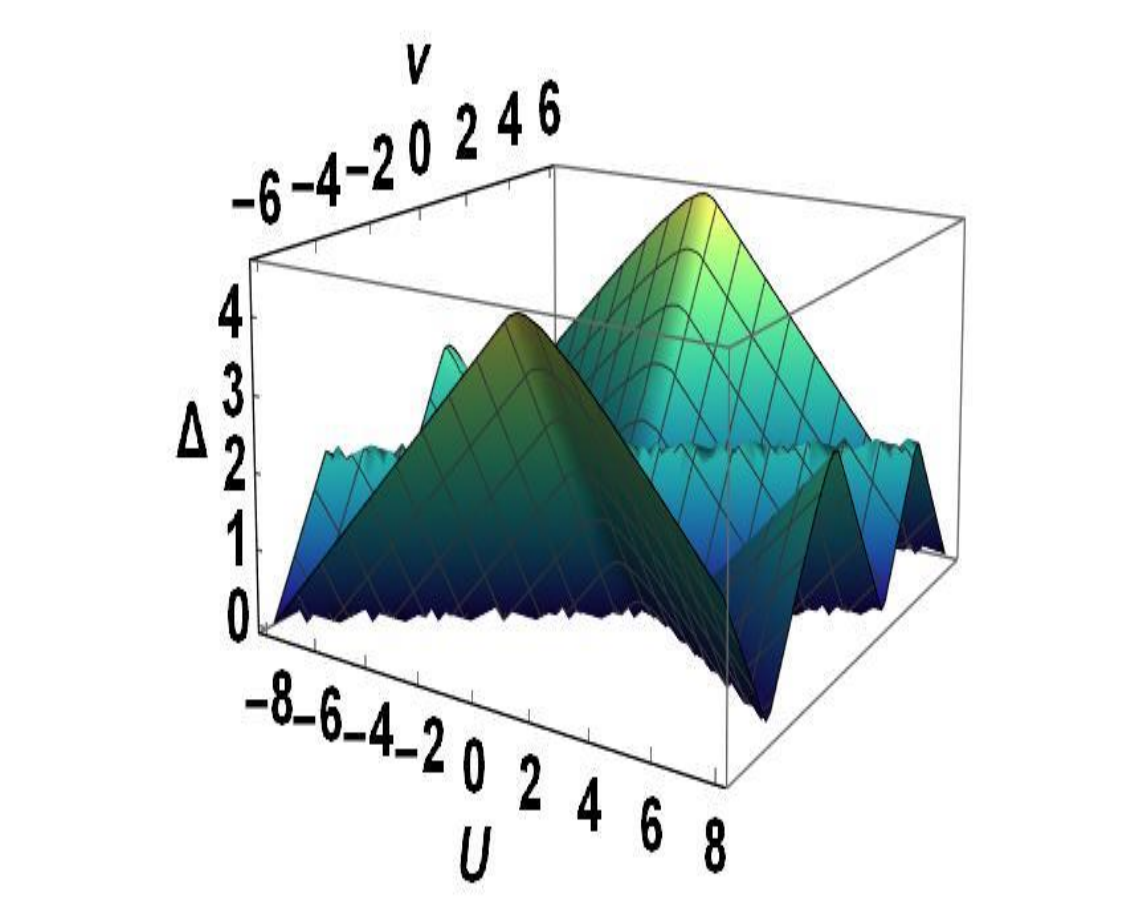} 
                  }
    \end{minipage}
    \caption{(Color online)
Gap magnitudes $\Delta$ as a function of $v$ and $U$.
}
\label{fig:4}
\end{figure}

The Hamiltonian (\ref{eq:1})-(\ref{eq:4}) defines a family of two
parametric solvable models, which reduce to the MN model at $v=0$ and
to the Kitaev chain with zero chemical potential at $v=0$ and $U=0$.
\begin{figure}[tp]
     \centering{\leavevmode}
     \begin{minipage}[h]{.5\linewidth}
\center{
\includegraphics[width=\linewidth]{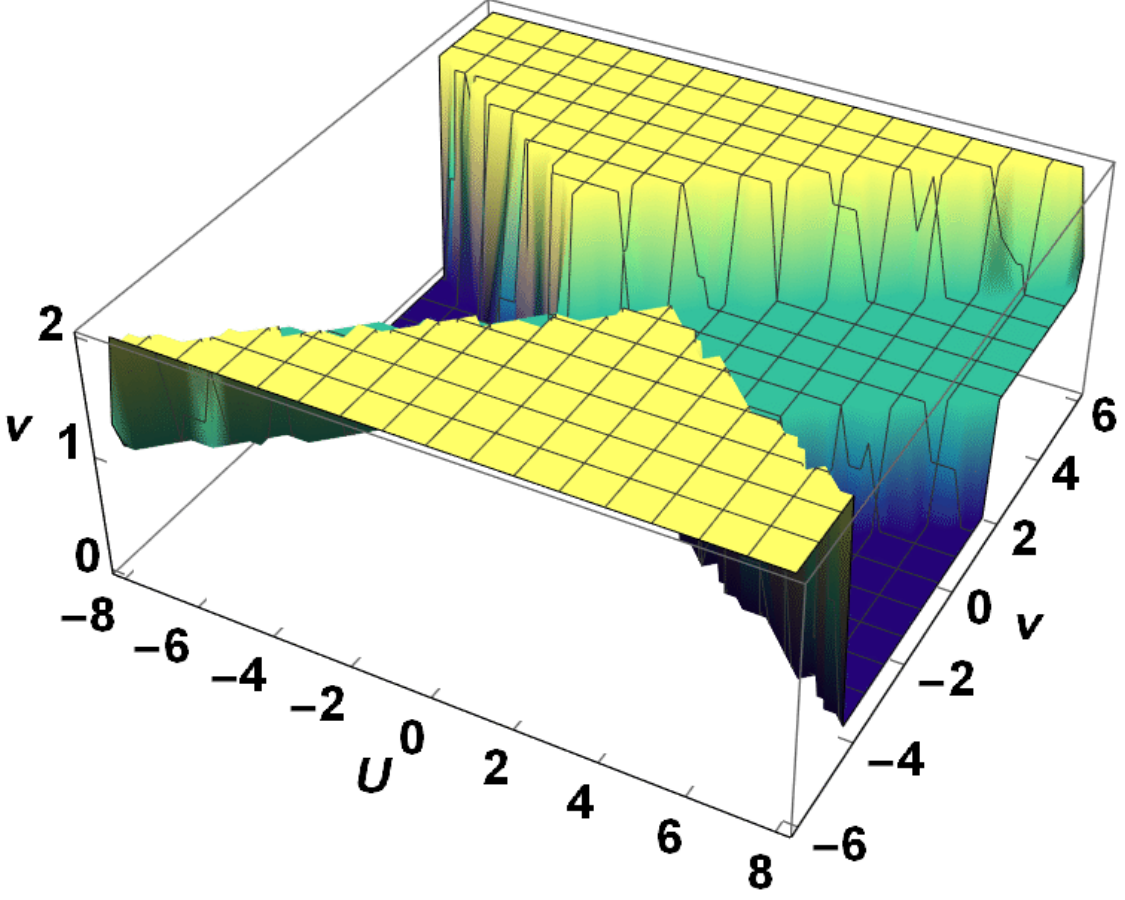} a)\\
                  }
    \end{minipage} \hfill
\begin{minipage}[h]{.45\linewidth}
\center{
\includegraphics[width=\linewidth]{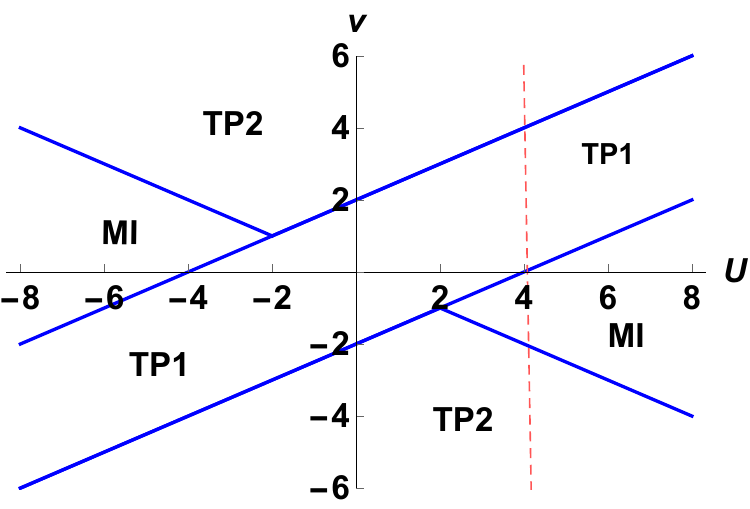} b)
                  }
    \end{minipage}
\caption{(Color online)
  The winding number invariant $\nu$ as a function of $U$ and $v$ a), $U-v$
  topological phase diagram b), curves $v=-\frac{1}{2}U$ at $\vert U \vert>2$
  and $v=\pm 2 +\frac{U}{2}$ separate the topological phases
  TP1 ($\nu =1$), TP2 ($\nu=2$), and MI ($\nu=0$). The dotted line
  marks the slice of the phase diagram at $U=4$, for which the spectrum of
  excitations is presented in Fig~4.
}
\label{fig:1}
\end{figure}

\begin{figure}[tp]
     \centering{\leavevmode}
\begin{minipage}[h]{.4825\linewidth}
\center{
\includegraphics[width=\linewidth]{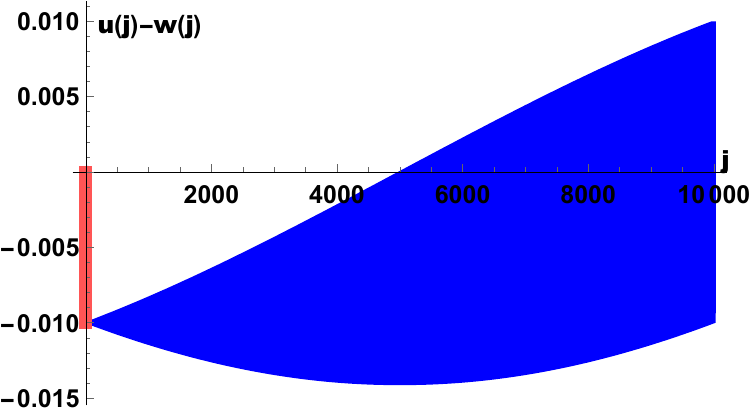} a)\\
                  }
    \end{minipage}
     \centering{\leavevmode}
\begin{minipage}[h]{.4825\linewidth}
\center{
\includegraphics[width=\linewidth]{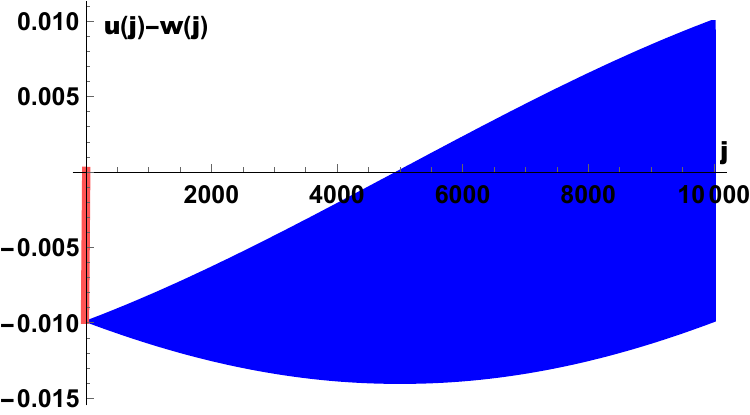} b)
                  }
    \end{minipage}
\caption{(Color online)
 Zero-energy Majorana edge wave functions $u(j)-w(j)$ as a function of
the coordinate along the chain $j$ ($L=10^4$) calculated at
TP1-TP2 phase transition point for $U=0$ and $v=2$. There are two
zero-energy Majorana edge wave functions localized at the boundaries
and two 'quasi' Majorana (partially localized at the boundary marked
in red) zero-energy edge wave functions with energies 0.00063 a) and
-0.00063 b).}
\label{fig:2}
\end{figure}
\begin{figure}[tp]
     \centering{\leavevmode}
    \begin{minipage}[h]{.65\linewidth}
\center{
\includegraphics[width=\linewidth]{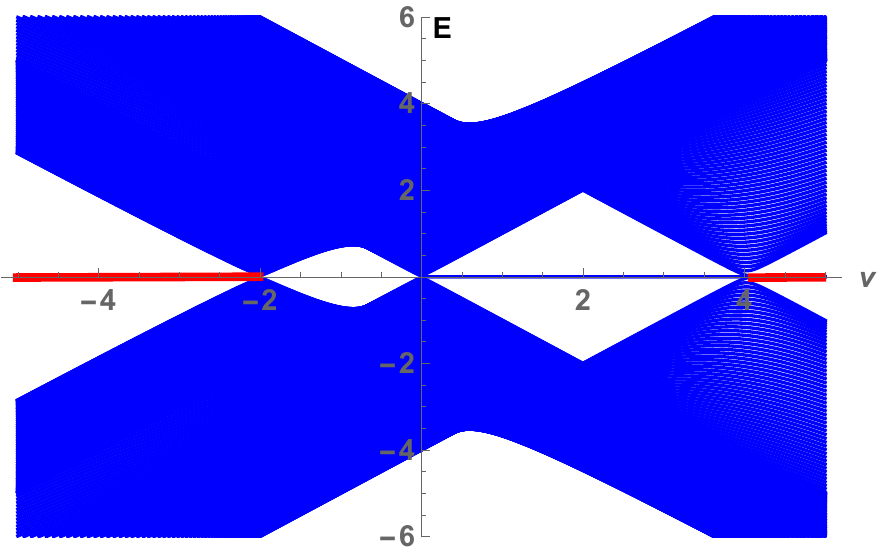} 
                  }
    \end{minipage}
    \caption{(Color online)
Energy band and edge states energy as a function of $v$ for $U=4$; at
$v<-2$ and $v>4$ TP2 phase is realized (with quadruply degenerate
  zero energy modes shown by red horizontal lines). A small pocket at $-2<v<0$
  corresponds to MI phase (without zero energy states), and a large pocket at
$0<U<4$ corresponds to TP1 phase (with doubly degenerate
zero energy modes shown by a blue horizontal line).
}
\label{fig:3}
\end{figure}

\section{Ground state phase diagram}
At half-filling, an important feature of the model is realized, which
concerns the topological states defining the phase diagram. The spectrum of
single-particle excitations is symmetric about the zero-point energy. It consists
of two branches separated by a gap in the charge excitation spectrum and has the
following form:
\begin{eqnarray}
&&E_{\pm}(k)=\pm \sqrt{4 t^2 + \frac{1}{4}U^2 +  v^2 - 2 t (U - 2 v) \cos k -  U v \cos 2k},
\label{eq:5}
\end{eqnarray}
where $k$ is the electron wave vector, without loss of generality it is convenient to put $t=1$.

The width of the gap in the charge excitation spectrum~(5) depends on
  the parameters of the Hamiltonian~(1) (see Fig.2). For any value of $U$ the bulk energy
  gap closes at $v=\pm 2 +\frac{U}{2}$, and for $\vert U \vert>2$ also at
  $v=-\frac{1}{2}U$. The disappearance of the gap coincides with the points of
  phase transitions between the trivial and nontrivial phases, which are
  characterized by the winding number invariant.
  
The MN approach reduces the
  Hubbard interaction to the motion of electrons in a $Z_2$-field. Its
  free configuration corresponds to the minimum energy of the system,
  so the Hamiltonian~(1) is reduced to a quadratic one, Eq.~(4). The
  approach proposed in [10] allows us to take into account the
  interaction between electrons, reducing the model Hamiltonian~(1) to
  a single-particle one. This makes it possible to use the well-known
  representation for the winding number invariant.
The topological states of the electron liquid is described by the winding number invariant, which is given by 
\begin{eqnarray}
\nu = \frac{1}{\pi}\int_{-\pi}^{\pi} dk \partial_k \varphi(k),
\label{eq:6}
\end{eqnarray}
where $\varphi(k)=\frac{1}{2}\arctan \left({\dfrac{2 \sin k +v \sin 2k }{-U/2+2 \cos k +v\cos 2k}}\right)$.

Numerical calculation of the winding number invariant as a function of $U$ and $v$ is shown in Fig 1a).
The MN criterion  $U_{MN} =\pm 4$, which corresponds to the point of phase transition between TP1 and MI phases, follows from  the condition $v=0$. In the phase diagram of Fig 1 b), the phases are separated by the phase boundaries $v=-\frac{1}{2}U$, at $\vert U \vert>2$ and $v=\pm 2 +\frac{U}{2}$. The condition $\nu=1$ corresponds to the topological phase TP1, in which there are two zero-energy Majorana fermions localized at the opposite boundaries of the sample, at $j=1$ and $j=L$. Zero-energy Majorana edge states correspond to different Majorana fermions, $\gamma_1=a_1+a^\dagger_1$ and $\chi_L=(a_L-a^\dagger_L)/i$, which are defined through the $a$-fermions.

This phase is realized in the Kitaev chain. Due to the correlated hopping of electrons between next-nearest neighbor sites new topological phase with $\nu=2$, denoted by TP2, is realized  at $\vert v \vert>\frac{1}{2}\vert U \vert$ and $v>2+\frac{U}{2}$ or $v<-2+\frac{U}{2}$, see Figs.~3a) and 3b). Two zero-energy Majorana edge modes are localized at each boundary, see Fig.~4). The phase diagram contains two critical points, $U=2,v=-1$ and $U=-2,v=1$, at which 
the three phases coexist. At the TP1-TP2 phase boundary, the edge modes that correspond to different topological phases coexist, see Fig.~4) Note that the two edge modes are partially localized. In Fig.~5, we show the electron excitation spectrum as a function of $v$ to better understand the behavior of the electron liquid in different phases. Below we present analytical calculations that confirm the above numerical results.  We will use the approach proposed in \cite{A5} for the investigation of zero energy modes in the Kitaev chain and in \cite{IK1} for the persistent current in the 2D topological superconductor.

\section{Conclusion}
 
An exactly solvable 1D model of interacting electrons is proposed, which takes into account the on-site Hubbard interaction and correlated hoppings between electrons located at the next-nearest neighbor sites. The ground state phase diagram of the model is rich, it includes two topological phases with different winding number invariants and a trivial topological phase of the Mott insulator. In the topological phases, zero-energy Majorana edge modes and winding number invariant determine the topological state of the electron liquid. Taking into account open boundary conditions,  eigenvalues and eigenvectors of the model Hamiltonian have been obtained and  illustrated by numerical calculations.

Our exact analytical results for a finite modified Kitaev-Hubbard chain provide a new insight into the behavior of the electron liquid in the topological and MI phases. We have delved into their nature and elucidated that there is a critical value of the amplitude repulsion interaction between fermions, which leads to the transition from the topological to MI phase of the electron liquid. The results are obtained within the framework of 1D exactly solvable models and contribute to the understanding of the topological to MI transition.

 \section{Appendix A. Eigenstate of a finite chain}
We start with the equation for the two-component wave function $u(j)$, $w(j)$ for the chain of length $L$, where $L$ is even and $-\frac{L}{2}\leq j \leq\frac{L}{2}$:
\begin{eqnarray}
&&(E-\frac{U}{2}) u(j)= - u(j+1)-u(j-1)-w(j+1)+w(j-1)-\nonumber \\&&
\frac{v}{2}[u(j+2)+u(j-2)]-\frac{v}{2}[w(j+2)-w(j-2)], \nonumber \\
&&(E+\frac{U}{2}) w(j)=  w(j+1)+w(j-1)+u_(j+1)-u(j-1)+\nonumber \\&&
\frac{v}{2}[w(j+2)+w(j-2)]+\frac{v}{2}[u(j+2)-u(j-2)],
\label{eq:A1}
\end{eqnarray}
for $-\frac{L}{2}+1< j <\frac{L}{2}-1$.
Equations for the wave function defined for the end points ($\pm \frac{L}{2},\pm (\frac{L}{2}-1)$) are valid at additional boundary conditions:
\begin{eqnarray}
&&\left(E-\frac{U}{2}\right) u\left(-\frac{L}{2}\right)= -  u\left(-\frac{L}{2}+1\right)- w\left(-\frac{L}{2}+1\right)-\frac{v}{2} u\left(-\frac{L}{2}+2\right)-\frac{v}{2} w\left(-\frac{L}{2}+2\right), \nonumber \\
&&\left(E+\frac{U}{2}\right) w\left(-\frac{L}{2}\right)=   w\left(-\frac{L}{2}+1\right)+ u\left(-\frac{L}{2}+1\right)+\frac{v}{2} w\left(-\frac{L}{2}+2\right)+\frac{v}{2} u\left(-\frac{L}{2}+2\right), 
 \nonumber \\&&
\textrm{at}
\nonumber\\&&u\left(-\frac{L}{2}-1\right)-w\left(-\frac{L}{2}-1\right)+\frac{v}{2}u\left(-\frac{L}{2}-2\right)-\frac{v}{2} w\left(-\frac{L}{2}-2\right)=0,
\label{eq:A2} 
\end{eqnarray}
\begin{eqnarray}
&&\left(E-\frac{U}{2}\right) u\left(-\frac{L}{2}+1\right)= -  u\left(-\frac{L}{2}+2\right)-u\left(-\frac{L}{2}\right)- w\left(-\frac{L}{2}+2\right)+w\left(-\frac{L}{2}\right)- \nonumber \\&&
\frac{v}{2} u\left(-\frac{L}{2}+3\right)-\frac{v}{2} w\left(-\frac{L}{2}+3\right), \nonumber \\
&&\left(E+\frac{U}{2}\right) w\left(-\frac{L}{2}+1\right)=  w\left(-\frac{L}{2}+2\right)+w\left(-\frac{L}{2}\right)+ u\left(-\frac{L}{2}+1\right)-u\left(-\frac{L}{2}\right)+ \nonumber \\
&&\frac{v}{2} w\left(-\frac{L}{2}+3\right)+\frac{v}{2} u\left(-\frac{L}{2}+3\right), \nonumber \\&&  \textrm{at} \nonumber\\&&
u\left(-\frac{L}{2}-1\right)-w\left(-\frac{L}{2}-1\right)=0,
\label{eq:A3}
\end{eqnarray}
\begin{eqnarray}
&&\left(E-\frac{U}{2}\right) u\left(\frac{L}{2}\right)= -  u\left(\frac{L}{2}-1\right)+ w\left(\frac{L}{2}-1\right)-\frac{v}{2} u\left(\frac{L}{2}-2\right)+\frac{v}{2} w\left(\frac{L}{2}-2\right), \nonumber \\
&&\left(E+\frac{U}{2}\right) w\left(\frac{L}{2}\right)=   w\left(\frac{L}{2}-1\right)- u\left(\frac{L}{2}-1\right)+\frac{v}{2} w\left(\frac{L}{2}-2\right)+\frac{v}{2} u\left(\frac{L}{2}-2\right),  \nonumber \\&&  \textrm{at} \nonumber\\&&
u\left(\frac{L}{2}+1\right)+w\left(\frac{L}{2}+1\right)+\frac{v}{2}u\left(\frac{L}{2}+2\right)+ \frac{v}{2}w\left(\frac{L}{2}+2\right)=0,
\label{eq:A4}
\end{eqnarray}
\begin{eqnarray}
&&\left(E-\frac{U}{2}\right) u\left(\frac{L}{2}-1\right)= - u\left(\frac{L}{2}-2\right)-u\left(\frac{L}{2}\right)- w\left(\frac{L}{2}\right)+w\left(\frac{L}{2}-2\right)-\nonumber \\
&&\frac{v}{2} u\left(\frac{L}{2}-3\right)+\frac{v}{2} w\left(\frac{L}{2}-3\right), \nonumber \\
&&
\left(E+\frac{U}{2}\right) w\left(\frac{L}{2}-1\right)=  w\left(\frac{L}{2}-2\right)+w\left(-\frac{L}{2}\right)+ u\left(\frac{L}{2}\right)-u\left(\frac{L}{2}-2\right)+\nonumber \\
&&\frac{v}{2} w\left(\frac{L}{2}-3\right)+\frac{v}{2} u\left(\frac{L}{2}-3\right),  \nonumber \\&& \textrm{at} \nonumber\\&&
u\left(\frac{L}{2}+1\right)+w\left(\frac{L}{2}+1\right)=0,
\label{eq:A5}
\end{eqnarray}
The expression Eq.~(\ref{eq:5}) for the energy $E$ follows from the solution of Eqs.~(\ref{eq:A1}) for the amplitudes $u(j)= u(k)\exp (i kj)$ and $w(j)=w(k)\exp(i k j)$, where $k$ is the electron momentum. Taking into account that the solution for the wave function exists for two exponentials with wave vectors $k$ and $-k$, it is convenient to define the amplitudes $u(k)$ and $w(k)$ in the following form:
\begin{eqnarray}
&&u(k)=A\cos(\varphi)\sin (k j +\varphi), \quad w(k)=A \sin(\varphi)\cos (k j +\varphi),
\label{eq:A6}
\end{eqnarray}
where $A$ is the the normalization factor.

Taking into account the solution for the wave function (\ref{eq:A6}), the values of $k$ are determined by the boundary conditions:
\begin{eqnarray}&&
 u(-L/2-2)-w(-L/2-2)=0  \to  \sin [k(L/2+2)]=0, \nonumber \\&&
 u(-L/2-1)-w(-L/2-1)=0  \to  \sin [k(L/2+1)]=0, 
 \label{eq:A7}
\end{eqnarray}
 \begin{eqnarray}&&
u(L/2+2)+w(L/2+2)=0  \to  \sin [k(L/2+2)+2\varphi]=0,\nonumber \\&&
 u(L/2+1)+w(L/2+1)=0  \to  \sin [k(L/2+1)+2\varphi]=0.
\label{eq:A8}
\end{eqnarray}
The solutions $k =\dfrac{2 \pi n}{L+2}, n=1,...,L+1$, Eq.~(\ref{eq:A7}), determine the bulk state of the electron liquid, and the other solutions,  Eq.~(\ref{eq:A8}), yield the edge states.

In the TP1 phase, zero energy edge states are realized for the momentum $k_1=\pi +i \delta$ at $U<4+2v$ and for $k_1=0 +i \delta$ at $U<-4+2v$. As a result, there is only one edge mode at the boundary, see Fig.~2a).
In the TP2 phase, zero-energy edge modes are realized for the momentum $k_{2}=\pm \arccos(\frac{1}{v})- i \delta$ at $U<-2v, \vert v \vert \geq 1$, so in this case we are talking about two edge modes at the same boundary, see Fig.~2b). According to Eq.~(\ref{eq:A8}), $\text{Im}\,\varphi=-\delta L/2$ and $\cos 2\varphi =\dfrac{U/2-2 \cos k -v \cos2k}{E}$. For $\delta L\gg 1$ zero energy corresponds to the edge modes. The solution for the wave function (\ref{eq:A6}) at the complex values of the momentum yields the edge states at $-\frac{L}{2}\leq j\leq \frac{L}{2}$, such that $u(j)-w(j)=A \sinh(\delta j)$ for $k_1=i \delta$  with $A^2=\dfrac{2\delta}{-\delta L +\sinh (\delta L)}$, where $k_1$ corresponds to the open boundary conditions (\ref{eq:A8}). At $\delta L\gg 1$ the wave function of the zero-energy edge state is localized  at the boundary, $u(j)-w(j)\simeq\sqrt {2\delta} \dfrac{\sinh( \delta j)}{\sqrt{\sinh^2(\delta L/2)}} \sim \exp [-\vert \delta \vert (L/2-j)]$.

\section{Appendix B. The 1D Falicov-Kimball model with $p$-pairing}

Let us consider a 1D model that describes the interaction between band and localized fermions, namely the Falicov-Kimball model
\begin{eqnarray}&&{\cal H}=-\frac{1}{2}\sum_{j}(t c^\dagger_{j+1}c_{j}+\Delta c_{j+1}c_{j} +H.c.)+ U\sum_{j}n_{j}m_{j}-
\epsilon \sum_{j}m_{j},
\label{eq:B1}\end{eqnarray}
where $c^\dagger_{j},c_j $and $d^\dagger_{j},d_j$ are the spinless fermion operators defined on a lattice site $\emph{j}$,  $U$ is the  value of the on-site interaction, $n_{j}=c^\dagger_{j}c_{j}-\frac{1}{2}$, $m_{j}=d^\dagger_{j}d_{j}-\frac{1}{2}$, $\epsilon >0$ is the energy of localized $d$-fermions. The Hamiltonian is a 1D Falicov-Kimball model in which the pairing of fermions $\Delta c_{j}c_{j+1}$  at the special point $\Delta =t$
 is taken into account.

In the spin representation for $c$- and $d$-operators via $\textbf{S}_j$ and $\textbf{T}_j$ spin-$\frac{1}{2}$ operators the Hamiltonian (\ref{eq:B1}) has the following form 
\begin{eqnarray}
&&{\cal H}= -2t\sum_{j}S_{j+1}^x S_{j}^x +U\sum_{j=1}^{N}S_j^zT_j^z-\epsilon\sum_{j}T^z_j=\nonumber\\&&
-2t\sum_{j}J_{j+1}^x J_{j}^x+\frac{U}{2}\sum_{j=1}^{N}J^z_j-\epsilon\sum_{j}P^x_j.
\label{eq:B2} 
\end{eqnarray}
Because the operator $P^x_j$ commutes with the Hamiltonian the last summand in Eq.~(\ref{eq:B2}) is a $c$-number. At the special point $t=\Delta$ the Hamiltonian (\ref{eq:B2}) coincides with the one of the MN model, see Eq.~(\ref{eq:3}), with $P^z_j=0$ and $v=0$. Thus, TP1 phase is stable at $\vert U\vert<2$, and the phase transition points $U=\pm 2 t$ corresponds to TP1-M1 phase transition.

\section*{Acknowledgments}
This work was supported in part by the Spanish Ministry of Science and Innovation,
Projects No.~PID2022-139230NB-I00 and No.~PID2022-138750NB-C22.


\end{document}